\documentclass{article}

\usepackage{arxiv}

\usepackage[utf8]{inputenc}
\usepackage[T1]{fontenc}
\usepackage{amsmath}
\usepackage{amsfonts}
\usepackage{amssymb}
\usepackage{amsthm}
\usepackage{booktabs}
\usepackage{nicefrac}      % optional
\usepackage{microtype}

\newtheorem{theorem}{Theorem}
\newtheorem{lemma}[theorem]{Lemma}
\newtheorem{proposition}[theorem]{Proposition}
\newtheorem{corollary}[theorem]{Corollary}
\theoremstyle{definition}
\newtheorem{definition}[theorem]{Definition}
\newtheorem{assumption}[theorem]{Assumption}
\newtheorem{requirement}[theorem]{Requirement}
\theoremstyle{remark}
\newtheorem{remark}[theorem]{Remark}
\newtheorem{example}[theorem]{Example}

\usepackage[round]{natbib}

\usepackage{hyperref}
\usepackage{url}
\date{September 2026}

\title{A Characterized Softmax for Distributing Standing Across Multiple Affiliations}

\author{
 Md Mahedi Hasan Riday \\
  University of Siegen\\
  Siegen, Germany \\
  \texttt{md.riday@student.uni-siegen.de} \\
  \And
  Md Shajalal \\
  University of Siegen\\
  Siegen, Germany \\
  \texttt{md.shajalal@uni-siegen.de} \\
  \And
  Gunnar Stevens \\
  University of Siegen\\
  Siegen, Germany \\
  \texttt{gunnar.stevens@uni-siegen.de} \\
}

\begin{document}
\maketitle
\begin{abstract}
Cooperative game theory answers a small number of canonical questions: payoff allocation (Shapley, Owen), interaction attribution (Grabisch--Roubens, Faith-Shap), and coalition formation and stability (the core, cooperative games with overlapping coalitions). None answers a fourth question that arises whenever players carry known, overlapping group tags fixed independently of the game: how should a player's standing be distributed across the several groups it already belongs to? We call this membership attribution, formalize it precisely, and give a complete characterization of its solution. The direct, ratio-based answer is provably invalid whenever a player's group-level signals have mixed sign --- a common, structurally unavoidable case, not a contrived one. We propose a restricted-domain softmax construction instead, and prove that, for any player satisfying a precisely defined richness condition on the group structure, it is the unique such map, up to a temperature parameter, satisfying five requirements. A sixth requirement fixes the sign of this temperature; a seventh, shown independent of the rest via an explicit counterexample, forces one shared temperature across all eligible players. We illustrate the construction on an exact cooperative game, confirming
every applicable requirement numerically --- including validity, scale invariance, and monotonicity, with symmetry and locality verified on separate minimal instances --- and showing it degrades gracefully, valid but uninformative, never invalid, whenever a player's own evidence fails to distinguish between its groups. The result is a closed-form rule, uniquely characterized under a stated compositional principle for separable interaction evidence, for distributing overlapping group membership, applicable wherever cooperative-game players carry known, overlapping tags: legislative committees, interdisciplinary research groups, or multi-unit organizations.
\end{abstract}

% keywords can be removed
%\keywords{First keyword \and Second keyword \and More}

\section{Introduction}
\label{sec:intro}

Cooperative game theory's existing tools cluster into three families, and it
is worth naming them explicitly. \textbf{Payoff allocation} asks how to split
a coalition's achieved value among its members --- Shapley's value
\citep{shapley1953}, and Owen's extension to a priori coalition structures
\citep{owen1977}, both answer this, but Owen's extension requires every
player to belong to exactly one structural unit. \textbf{Interaction
attribution} asks which subsets of players interact, and by how much ---
Grabisch--Roubens' interaction indices \citep{grabisch1999} and Faith-Shap
\citep{tsai2023faithshap} answer this, generalizing the pairwise Harsanyi
dividend to arbitrary subset size, but presuppose no exogenous group
structure exists at all. \textbf{Coalition formation and stability} asks
which coalition structures can form and persist --- cooperative games with
overlapping coalitions \citep{chalkiadakis2010,zick2019} answer this, but
treat a player's fractional participation as a decision variable to be found
or verified under an equilibrium or optimality criterion, an inherently
algorithmic question, rather than a statistic read off a fixed game's own
interaction structure.

None of these three families answers a fourth question that arises whenever
players carry known, overlapping group tags fixed \emph{independently} of the
game being played --- a legislator on two committees, a firm's division
reporting to two business units. Given that a player's affiliations are
already known and fixed, and the game's own interaction structure is already
computed, how should that player's standing be distributed across its
several known groups? This is not payoff allocation (the groups, not the
payoff split, are exogenous); not interaction attribution (the group
taxonomy is given, not derived from which subsets interact); and not
coalition formation (nothing is being formed or stabilized --- the groups
already exist). We call this \emph{membership attribution}, and it is the
problem this paper formalizes and completely solves.

\textbf{The naive approach fails.} The most direct way to convert a player's net interaction with each of its known groups into a normalized weight is to compute a raw signal per group and divide by their sum. This fails whenever a player's groups combine both synergistic and redundant interactions --- a structurally unavoidable case once pairwise interaction terms can carry either sign, which they always can in a general cooperative game (Section~\ref{sec:construction}).

\textbf{Contributions.}
\begin{enumerate}
\item We identify membership attribution as a distinct question cooperative
game theory has not previously formalized (Section~\ref{sec:related}), and
give it a precise mathematical statement (Section~\ref{sec:construction}).
\item We prove the direct, ratio-based answer to this question is impossible on any profile mixing synergy and redundancy --- a common, structurally unavoidable case, not a corner case.
\item We give a construction --- a restricted-domain softmax over a
player's known groups --- and prove it is the unique solution satisfying
five requirements, under a precisely defined richness condition, with a
sixth requirement shown to pin down the sign of the construction's free
parameter, and a seventh, shown logically independent of the rest via an
explicit, fully verified counterexample, forcing a single shared
temperature across players (Sections~\ref{sec:axioms}--\ref{sec:uniqueness}).
\item We identify the construction's one genuine degenerate case and show it
degrades gracefully rather than silently (Section~\ref{sec:discussion}).
\end{enumerate}

\textbf{Roadmap.} Section~\ref{sec:related} places this contribution against
prior extensions of cooperative game theory. Section~\ref{sec:prelim} builds
the mathematical foundation. Section~\ref{sec:construction} states the
problem, proves why direct normalization fails, and derives the
construction. Section~\ref{sec:axioms} states and proves six requirements.
Section~\ref{sec:uniqueness} defines richness precisely and proves
uniqueness, including the sign of the temperature parameter.
Section~\ref{sec:empirical} works through the construction numerically on an exact cooperative game. Section~\ref{sec:discussion} discusses limitations
and future work. Section~\ref{sec:conclusion} concludes.

%%%%%%%%%%%%%%%%%%%%%%%%%%%%%%%%%%%%%%%%%%%%%%%%%%%%%%%%%%%%%%%%%%%%%%%%

\section{Related Work}
\label{sec:related}

We organize prior work by the three question-families introduced in Section~\ref{sec:intro}, then by the mathematical machinery our uniqueness
proof draws on.

\textbf{Payoff allocation.} \citeauthor{owen1977} (\citeyear{owen1977})
generalizes the Shapley value to games with an a priori coalition structure:
a partition of the player set into disjoint unions. This assumes every player
belongs to exactly one union; it cannot represent a player belonging to
several groups simultaneously. The Harsanyi set of payoff vectors
\citep{vasil2001harsanyi} generalizes \emph{how} a single coalition's own dividend
is split among its members, but still presupposes each player sits inside
one coalition being evaluated at a time, not several fixed, overlapping
groups at once.

\textbf{Interaction attribution.} \citet{grabisch1999} axiomatize interaction
indices among arbitrary coalitions. Faith-Shap \citep{tsai2023faithshap}
characterizes an interaction index as the coefficient of the most faithful
polynomial approximation to a value function. Both answer which subsets of
players interact, and by how much --- presupposing no group structure exists
at all.

\textbf{Coalition formation and stability.} OCF games
\citep{chalkiadakis2010,zick2019} generalize the domain of the characteristic
function itself to fractional participation profiles, studying how to find
or verify optimal or stable such profiles --- shown NP-hard in general,
tractable only under structural restrictions. The framework remains an
active one: recent applications model overlapping coalition formation among
resource-constrained agents in edge-computing networks
\citep{ye2025overlapping} and among suppliers in electricity markets
\citep{valizadeh2024enhancing}, both within the same decision-variable framing. The
fractional participation level there is a decision variable in an
optimization or equilibrium problem. Our construction keeps $v$ an ordinary
crisp-coalition game and returns a closed-form statistic --- there is no
search, no stability notion, no optimality criterion.

\textbf{Related mathematical machinery.} Two further lines of work are
adjacent without answering our question. \citet{aubin1981} and
\citet{butnariu1980} extend the domain of $v$ to a graded participation
vector for a \emph{single} coalition being evaluated, not how one player's
membership splits across several, simultaneously held, discrete group
labels; combining a priori union structures with fuzzy coalitions
\citep{sun2017fuzzyowen} retains the single-union assumption and so
inherits Owen's limitation above. Separately, our uniqueness theorem rests
on Luce's Choice Axiom \citep{luce1959}, adapted here to a deterministic
setting and proved in a self-contained way rather than merely invoked, and
is consonant with \citeauthor{mcfadden1974}'s (\citeyear{mcfadden1974})
conditional logit model; neither addresses group membership at all, since
both concern choice among alternatives, not the division of a fixed
quantity across a chooser's own overlapping affiliations.

No prior construction, in any of these four lines of work, takes as input
both (i) a fixed game's own pairwise interaction structure and (ii) an
exogenous, possibly overlapping group membership map, and returns a
normalized fuzzy membership distribution in closed form --- the gap this
paper closes.

%%%%%%%%%%%%%%%%%%%%%%%%%%%%%%%%%%%%%%%%%%%%%%%%%%%%%%%%%%%%%%%%%%%%%%%%

\section{Preliminaries}
\label{sec:prelim}

This section sets up the two pieces of standard cooperative-game-theory
machinery the rest of the paper is built from: a way to measure how much
value a group of players creates together, and a way to isolate exactly how
much of that value comes from any one \emph{pair} of players helping or
hurting each other. Readers already familiar with cooperative games and
Harsanyi dividends can skip to Section~\ref{sec:construction}; everyone
else, the intuition below is worth reading slowly, since every later
section reuses these two ideas without redefining them.

\textbf{Cooperative games.} A cooperative game is just a rule for how much
value \emph{any} subset of players can produce together, without saying how
that value gets split among them --- the splitting question is handled
separately, elsewhere. Formally:

\begin{definition}[Cooperative game]
$(N,v)$: $N=\{1,\dots,n\}$ a finite player set, $v:2^N\to\mathbb{R}$ with
$v(\emptyset)=0$.
\end{definition}

So $v(S)$ is a single number --- the total value the players in $S$ create
when they work together --- and $v(\emptyset)=0$ just says a group of no
one creates nothing. Nothing here assumes the value function behaves nicely
(it need not be additive, monotonic, or anything else); the theory earns its
structure from what comes next.

\textbf{Isolating one subset's own contribution.} Knowing $v(S)$ for every
$S$ tells us the \emph{total} value any group can produce, but it does not
by itself tell us how much of that total is attributable to a specific
subset's members cooperating with each other, as opposed to simply
reflecting what its members would have produced on their own or in smaller
combinations. The Harsanyi dividend is the standard way to strip out that
overlap and isolate exactly what each subset adds on top of everything
smaller than it:

\begin{definition}[Harsanyi dividend]
For $A\subseteq N$, $\varphi_v(A)$ is the unique solution to
$v(S)=\sum_{A\subseteq S}\varphi_v(A)$ for every $S\subseteq N$.
\end{definition}

In words: the dividend $\varphi_v(A)$ is defined so that summing it over
every subset contained in $S$ exactly reconstructs $v(S)$. This is a system
of equations, one per subset $S$, and before using it we need to know it
actually has a solution, and only one. It does, and the next lemma also
hands us a direct formula instead of leaving $\varphi_v$ defined only
implicitly:

\begin{lemma}[Existence and uniqueness via Möbius inversion]
\label{lem:mobius}
$\varphi_v(A)=\sum_{B\subseteq A}(-1)^{|A|-|B|}v(B)$, where $B$ ranges over
subsets of $A$ as a summation index.
\end{lemma}

\begin{proof}
Möbius inversion on $(2^N,\subseteq)$: for any $S$,
$\sum_{A\subseteq S}\varphi_v(A)=\sum_{B\subseteq S}v(B)(1-1)^{|S\setminus B|}$,
which equals $v(S)$ since $(1-1)^m=1$ iff $m=0$.Uniqueness follows because the system is triangular: each equation for
$\varphi_v(A)$ involves $v(S)$ together only with $\varphi_v(B)$ for
$B\subsetneq A$, already determined by strictly smaller subsets, so
solving in order of increasing subset size determines every value in
turn, with nothing left underdetermined or overdetermined. \qed
\end{proof}

The alternating-sign sum is an instance of inclusion-exclusion: it adds back
and subtracts off smaller subsets' values until all that is left is what
$A$ itself contributes once everything smaller has already been accounted
for. The only case this paper actually needs is the smallest nontrivial
one, a \emph{pair} of players:
$$\varphi_v(\{i,j\})=v(\{i,j\})-v(\{i\})-v(\{j\}).$$
Read directly, this says: take what $i$ and $j$ produce together, and
subtract what each produces alone. Whatever is left over cannot be
explained by either player individually --- it is specifically what their
\emph{pairing} adds or removes. This quantity is important enough to this
paper to name:

\begin{definition}[Pairwise interaction term]
\begin{equation}
\phi_{ij}:=v(\{i,j\})-v(\{i\})-v(\{j\}) = \varphi_v(\{i,j\})
\label{eq:pairwise}
\end{equation}
\end{definition}

The pairwise interaction term is not a new primitive; it is the Harsanyi
dividend of the two-player coalition (Lemma~\ref{lem:mobius} with $A=\{i,j\}$,
using $v(\emptyset)=0$), so every property proved of Harsanyi dividends
applies to it, including $\phi_{ij}=\phi_{ji}$ --- how much $i$ and $j$ help
or hurt each other does not depend on which one we list first.

The sign of $\phi_{ij}$ is what the rest of the paper is organized around,
so it is worth stating plainly rather than leaving it implicit:

\begin{remark}
\label{rem:sign}
$\phi_{ij}>0$: synergy; $\phi_{ij}<0$: redundancy; $\phi_{ij}=0$:
independence. This sign is forced by \eqref{eq:pairwise}, not chosen.
\end{remark}

Concretely: if $i$ and $j$ together are worth more than the sum of what
each is worth alone, they are \emph{synergistic} and $\phi_{ij}>0$; if
together they are worth less --- because, say, they duplicate each other's
contribution --- they are \emph{redundant} and $\phi_{ij}<0$; and if
pairing them changes nothing, $\phi_{ij}=0$. Nothing about the construction
in this paper chooses this sign or restricts it in advance: it is read off
whatever cooperative game is given, and it can go either way for different
pairs in the same game. That single fact --- that $\phi_{ij}$ is a real
number of \emph{either} sign, not a nonnegative quantity --- is the root
cause of why the naive approach to membership attribution fails
(Section~\ref{sec:construction}), and everything built in
Sections~\ref{sec:construction}--\ref{sec:uniqueness} exists to handle it
correctly.

%%%%%%%%%%%%%%%%%%%%%%%%%%%%%%%%%%%%%%%%%%%%%%%%%%%%%%%%%%%%%%%%%%%%%%%%

\section{Problem Statement and Construction}
\label{sec:construction}

Section~\ref{sec:prelim} gave us $\phi_{ij}$, a signed number measuring how
much any two players help or hurt each other. This section uses that
signal to state the paper's actual problem precisely, shows why the most
obvious way to solve it is mathematically broken, and then builds a
construction that is not.

\textbf{Group affiliations are given, not derived.} The groups a player
belongs to --- which committees a legislator sits on, which business units
a division reports to --- are assumed to be already known before we look
at the game at all. The game tells us how players interact; it says
nothing about which groups they belong to, and this paper never tries to
infer group membership from $\phi_{ij}$. The notation below just names
this externally-given structure:

\begin{definition}[Known group affiliations]
For $i\in N$, $\mathcal{G}_i\subseteq G$ is the known, exogenous set of
groups $i$ belongs to. $M_g:=\{k\in N:g\in\mathcal{G}_k\}$.
\end{definition}

So $\mathcal{G}_i$ is simply the list of groups player $i$ is in, and
$M_g$ runs the lookup the other way: given a group, who is in it. A player can appear in $\mathcal{G}_i$ for several different $g$ at once --- that is precisely the overlap this paper is about --- and a player with no known groups at all would make the problem vacuous, so we rule that out. To be
precise about what ''membership attribution'' denotes: we do not infer or quantify uncertainty about whether $i$ belongs to $g$; affiliations are observed and fixed in advance. The problem is to allocate a player's interaction-derived standing across its already-known memberships, not to determine what those memberships are.

\begin{assumption}[Nonempty affiliation]
$\mathcal{G}_i\neq\emptyset$ for every $i\in N$.
\end{assumption}

\textbf{What a valid answer must look like.} Before building any particular
rule, it is worth pinning down what we would even accept as a correct
output. Intuitively, splitting a player's standing across its groups
should behave like splitting a whole into percentages: every piece greater than 0\% and up to 100\%, the pieces for groups the player is actually in should add
up to the whole, and a player gets no credit at all in a group it does not
belong to.

\begin{definition}[Fuzzy overlapping group membership]
\label{def:fuzzy-membership}
For every nonempty sub-menu $S\subseteq\mathcal{G}_i$, a \emph{fuzzy
overlapping group membership} is a menu-indexed family $\{w_{i,g}^S\}$
satisfying $w_{i,g}^S\in(0,1]$ for $g\in S$ (M1), $\sum_{g\in
S}w_{i,g}^S=1$ (M2), $w_{i,g}^S=0$ for $g\notin S$ (M3). We write
$w_{i,g}:=w_{i,g}^{\mathcal{G}_i}$ for the full-menu case.
\end{definition}

Any rule we propose is judged against these three conditions on every
sub-menu: (M1) keeps every weight a legitimate fraction, (M2) makes the
weights for a given sub-menu add to a whole, and (M3) says a player cannot
be given credit in a group outside that sub-menu. Until
Section~\ref{sec:uniqueness} introduces comparisons across sub-menus,
Sections~\ref{sec:construction}--\ref{sec:axioms} only ever need the
full-menu case $w_{i,g}=w_{i,g}^{\mathcal{G}_i}$, so we use the simpler
notation $w_{i,g}$ throughout until then. Membership attribution, as this
paper defines it, is exactly the problem of producing such a family from
the game's own interaction structure.

\subsection{The Raw Signal}

The first natural step is to ask, for a given player $i$ and one of its
groups $g$: summed across everyone else who is also in $g$, does $i$ tend
to help or hurt that group? Adding up $i$'s pairwise interaction terms with
every other member of $g$ answers exactly that:

\begin{equation}
    c_{i,g}:=\sum_{k\neq i,\,g\in\mathcal{G}_k}\phi_{ik}.
    \label{eq:rawsignal}
\end{equation}

A large positive $c_{i,g}$ means $i$ is, on net, synergistic with group
$g$'s other members; a large negative $c_{i,g}$ means $i$ is, on net,
redundant with them; and nothing yet stops $c_{i,g}$ from landing on either
side, since it is just a sum of terms that can themselves be positive or
negative. Before using this particular sum, it is worth checking that it
is not an arbitrary choice --- that any reasonable way of combining a
player's interactions into one group-level number has to look like this,
up to a constant:

\begin{lemma}[Sum form, dimension-consistent]
\label{lem:sumform}
Suppose that for every $m\ge1$, $c^{(m)}:\mathbb{R}^m\to\mathbb{R}$ is
symmetric, additive, continuous, and non-degenerate
($c^{(m)}(\delta_k)>0$), and that the family $\{c^{(m)}\}$ is
\emph{consistent under padding}: for any $m'<m$ and $x\in\mathbb{R}^{m'}$,
embedding $x$ into $\mathbb{R}^m$ by appending zeros satisfies
$c^{(m)}(x,0,\dots,0)=c^{(m')}(x)$. Then there is a single $a>0$,
independent of $m$, with $c^{(m)}(x)=a\sum_k x_k$ for every $m$; up to this
one shared normalization, $c$ is formula \eqref{eq:rawsignal}.
\end{lemma}

\begin{proof}
Fix $m$ and $k$; $h(t):=c^{(m)}(t\delta_k)$ satisfies the additive Cauchy
equation and is continuous, hence linear ($h(t)=a_mt$; continuity at one
point excludes pathological, nowhere-continuous solutions that the
unconstrained Cauchy equation would otherwise admit). Additivity across
coordinates gives $c^{(m)}(x)=\sum_k a_m x_k$; symmetry equalizes the
coefficients; non-degeneracy forces $a_m>0$. It remains to show $a_m$ does
not depend on $m$: fix any $x_1\in\mathbb{R}$ and pad it into
$\mathbb{R}^m$ for arbitrary $m$; consistency under padding gives
$c^{(m)}(x_1,0,\dots,0)=c^{(1)}(x_1)$, i.e.\ $a_m x_1=a_1 x_1$ for every
real $x_1$, hence $a_m=a_1$ for every $m$. Set $a:=a_1$. \qed
\end{proof}

In plain terms: if we only ask that combining interaction terms treats
every co-member the same way (symmetric), that adding one more co-member's
term changes the total in the obvious additive way (additive), that small
changes in the inputs only cause small changes in the output
(continuous), that at least one single positive interaction should count
for something (non-degenerate), and that a co-member who contributes
nothing ($\phi_{ik}=0$) changes nothing about the group's signal (padding
consistency --- already true of plain summation itself), then plain
summation with one shared constant across every group size is forced ---
no other candidate satisfies these five assumptions, for any group size.
This justifies treating $c_{i,g}$ as \emph{the} raw signal under this
stated set of assumptions, rather than one arbitrary choice among many.

\subsection{Why Direct Normalization Fails}

With a raw, signed number $c_{i,g}$ in hand for each of player $i$'s
groups, the obvious next step is to turn these numbers into percentages
the way one normally would: divide each one by some total. In symbols,
consider any rule of the form

\begin{equation}
    w_{i,g}\overset{?}{=}\frac{c_{i,g}}{h(c_{i,\cdot})},\qquad
h:\mathbb{R}^{\mathcal{G}_i}\to\mathbb{R}
\label{eq:ratiorule}
\end{equation}
where $h$ turns the whole list of $i$'s raw signals into a single shared
denominator. The simplest choice, (NAIVE), just sums them:

\begin{equation*}
w_{i,g} \overset{?}{=} \frac{c_{i,g}}{\sum_{g'} c_{i,g'}}
\end{equation*}

This is exactly how one would compute an ordinary percentage breakdown,
and it works perfectly well when every $c_{i,g}$ happens to be positive.
The trouble is that $c_{i,g}$ is a signed number (remark~\ref{rem:sign}), and once a player is synergistic with one group but
redundant with another, dividing by a shared total stops making sense:
a negative numerator over a positive denominator gives a negative
``percentage,'' which (M1) does not allow. The next result shows this is
not a flaw specific to (NAIVE) --- \emph{no} choice of shared denominator
$h$ can fix it:

\begin{proposition}[No shared-denominator normalization rule of the form \eqref{eq:ratiorule} satisfies (M1) on a mixed-sign profile]
\label{prop:linear-ratio-fails}
Fix any player $i$ and profile with $g^+,g^-\in\mathcal{G}_i$,
$c_{i,g^+}>0>c_{i,g^-}$. Then for \emph{every} function $h$, rule \eqref{eq:ratiorule} violates (M1) on this profile.
\end{proposition}

\begin{proof}
Suppose $h^*:=h(c_{i,\cdot})$ makes $w_{i,g}=c_{i,g}/h^*\in[0,1]$ for every
$g$. At $g^+$: $c_{i,g^+}/h^*\ge0$ with $c_{i,g^+}>0$ forces $h^*>0$. At
$g^-$: $c_{i,g^-}/h^*\ge0$ with $c_{i,g^-}<0$ forces $h^*<0$. Contradiction: 
no real $h^*$, and hence no function $h$, works.
\end{proof}

The argument is short because the contradiction is structural: a single shared denominator $h^*$ would have to be simultaneously positive (to keep the positive-signal group's weight non-negative) and negative (to keep the negative-signal group's weight non-negative), which is impossible for any real number. This rules out every rule whose numerator is the raw signal $c_{i,g}$ divided by a single shared denominator, not merely (NAIVE), whenever a player's groups mix synergy and redundancy --- a common, structurally unavoidable case. It does not rule out normalization schemes built on a different numerator, such as $e^{\beta c_{i,g}}/\sum_{g'}e^{\beta
c_{i,g'}}$: Section~\ref{sec:construction} turns to exactly such a construction next, precisely because its numerator is no longer the raw signal itself.

\subsection{The Construction}

The fix is to stop dividing by a shared total built from the raw signed
signals themselves, and instead first convert each signal into something
that is \emph{always} positive before dividing. Exponentiating does exactly
this: $e^{\beta c_{i,g}}$ is a positive number no matter what sign
$c_{i,g}$ has, so dividing by the sum of these transformed, always-positive
quantities can never produce a negative weight.

\begin{definition}[Restricted-domain softmax]
For any nonempty sub-menu $S\subseteq\mathcal{G}_i$,
$$w_{i,g}^S:=\frac{e^{\beta c_{i,g}}}{\sum_{g'\in S}e^{\beta c_{i,g'}}}
\text{ for }g\in S,\qquad w_{i,g}^S:=0\text{ for }g\notin S,\qquad\beta>0.$$
Writing $w_{i,g}:=w_{i,g}^{\mathcal{G}_i}$ for the full-menu case gives the
form used throughout Sections~\ref{sec:construction}--\ref{sec:axioms}:
\begin{equation}
    \boxed{\ w_{i,g}=\frac{e^{\beta c_{i,g}}}{\sum_{g'\in\mathcal{G}_i}e^{\beta
c_{i,g'}}}\ },\ \beta>0;\quad w_{i,g}:=0\text{ for }g\notin\mathcal{G}_i.
\label{eq:softmax}
\end{equation}
\end{definition}

The parameter $\beta$ controls how sharply the construction reacts to
differences between a player's groups: larger $\beta$ pushes weight more
aggressively toward whichever group has the highest raw signal, while
$\beta$ near $0$ spreads weight almost evenly regardless of the signals.
(Section~\ref{sec:uniqueness} shows this parameter is not a free design
choice but is itself pinned down by the paper's requirements.) Formula~\eqref{eq:softmax} evades
Proposition~\ref{prop:linear-ratio-fails} precisely because its effective
numerator, $e^{\beta c_{i,g}}$, is always strictly positive regardless of
the sign of $c_{i,g}$. The next proposition confirms this construction
always satisfies the validity conditions (M1)--(M3) laid out earlier, with
no caveats or exceptions:

\begin{proposition}[Formula~\eqref{eq:softmax} is always valid]
\label{prop:validity}
For every real profile and every $\beta>0$: $w_{i,g}\in(0,1]$ for
$g\in\mathcal{G}_i$, $0$ otherwise, summing to 1; the weight is strictly
less than $1$ whenever $|\mathcal{G}_i|\ge2$, and equals $1$ exactly when
$\mathcal{G}_i=\{g\}$ (Remark~\ref{rem:degeneracy}).
\end{proposition}

\begin{proof}
$e^{\beta c_{i,g}}>0$; the denominator, a finite sum of such terms (at
least one, by the nonempty-affiliation assumption), is strictly positive.
A positive term over a positive sum containing it lies in $(0,1]$, with
equality only when it is the sum's sole term; summing gives 1. \qed
\end{proof}

So unlike every rule in the shared-denominator family
\eqref{eq:ratiorule}, this construction never breaks, regardless of how a
player's raw signals happen to be signed. What remains is to show this
particular construction is not just \emph{a} valid fix, but the \emph{only}
one a short list of natural requirements will allow --- that is the
subject of Sections~\ref{sec:axioms} and~\ref{sec:uniqueness}.

%%%%%%%%%%%%%%%%%%%%%%%%%%%%%%%%%%%%%%%%%%%%%%%%%%%%%%%%%%%%%%%%%%%%%%%%
\section{Requirements}
\label{sec:axioms}

Section~\ref{sec:construction} built one specific rule --- the
restricted-domain softmax --- and showed it always produces valid
output. But ``always valid'' does not mean ``the only reasonable choice'':
many other formulas could also avoid the failure in
Proposition~\ref{prop:linear-ratio-fails}. This section writes down, in
precise mathematical terms, a short list of properties that any sensible
membership-attribution rule ought to have, independent of whether that
rule happens to be the softmax. Each requirement below is stated as a
plain-language description of a desirable behavior, immediately followed by
its exact mathematical meaning, and then checked against
formula~\eqref{eq:softmax}. Requirements 1--6 are \emph{existence-flavored}:
they are properties a single, already-given rule can be checked against
directly, and this section verifies formula~\eqref{eq:softmax} satisfies
all of them. Requirements 7--10, introduced in
Section~\ref{sec:uniqueness}, are \emph{uniqueness-flavored}: rather than
checking one rule, they are used to argue that softmax is the \emph{only}
rule satisfying everything demanded of it.

\begin{requirement}[Symmetry]
\label{req:symmetry}
Two players with identical interactions with every
other player, and the same known groups, receive identical weights: If $\mathcal{G}_i=\mathcal{G}_j$ and
$\phi_{ik}=\phi_{jk}$ for all $k\notin\{i,j\}$, then $w_{i,g}=w_{j,g}$.
\end{requirement}

This is the most basic fairness demand possible: if two players are, in every respect the game can see, interchangeable --- same groups, same interaction with everyone else --- the rule should not be able to tell them apart. A rule that broke ties between identical players using nothing but listing order would be a serious defect.

\begin{proof}
$c_{i,g}=c_{j,g}$ term-by-term, hence $w_{i,g}=w_{j,g}$ by \eqref{eq:softmax}. \qed
\end{proof}

\begin{requirement}[Scale Invariance]
\label{req:scale}
Rescaling every interaction by a positive constant
is exactly compensated by an inversely rescaled temperature: For $\lambda>0$, the pair $(\phi,\beta)$ and
$(\lambda\phi,\beta/\lambda)$ produce identical weights.
\end{requirement}

In other words, the \emph{units} the interaction terms happen to be
measured in should not matter. If every $\phi_{ij}$ were doubled --- say
because value were reported in cents instead of dollars --- the resulting
weights should be recoverable exactly by halving $\beta$ to compensate.
Nothing about the shape of the output should hinge on an arbitrary choice
of measurement scale.

\begin{proof}
$c_{i,g}\to\lambda c_{i,g}$; with $\beta\to\beta/\lambda$, the exponent
$(\beta/\lambda)(\lambda c_{i,g})=\beta c_{i,g}$ is unchanged. \qed
\end{proof}

\begin{remark}
This is a closure property of the whole parametrized family
$\{\mathrm{softmax}_\beta\}_{\beta\in\mathbb{R}}$: a \emph{fixed}-$\beta$
rule is \emph{not} invariant under $\phi\to\lambda\phi$ alone. The property
holds identically for any rule whose temperature does not depend on
$\phi$'s scale at all, which is why it carries no discriminating power in
Section~\ref{sec:uniqueness}.
\end{remark}

\begin{requirement}[Monotonicity]
\label{req:mono}
Stronger synergy with a group's member increases weight towards that group, provided some other known group is left unaffected to compare against: Fix a player $i$, a group $g \in \mathcal{G}_i$, and $k \in M_g \setminus \{i\}$. Suppose some
$g^*\in\mathcal{G}_i\setminus\{g\}$ has $k\notin M_{g^*}\setminus\{i\}$.
Then, holding all other $\phi_{i\ell}$ ($\ell\ne k$) fixed, $w_{i,g}$ is a strictly
increasing function of $\phi_{ik}$ alone, with no restriction on sign.
\end{requirement}

This is the property that gives the whole construction its point: if a
player becomes more synergistic with one of its groups --- and nothing
else about the picture changes --- that group's weight should strictly go
up, not down or stay flat. The proviso (requiring some \emph{other} group
to be unaffected) exists only so there is something to compare against; it
rules out the degenerate case where every one of a player's groups is
pushed by exactly the same amount, in which case nothing about the
\emph{relative} split has actually changed (see the remark below).

\begin{proof}
\textbf{Case 1 ($k$ affects only $g$).} $\partial w_{i,g}/\partial c_{i,g}
=\beta w_{i,g}(1-w_{i,g})$, strictly positive for every real $c_{i,g}$
\emph{because} $\beta>0$ (Definition of \eqref{eq:softmax}) and $w_{i,g}\in(0,1)$
(Proposition~\ref{prop:validity}). This is the only place in this paper
where $\beta>0$ is used to \emph{derive} a consequence rather than assumed
by fiat --- it is what lets this requirement pin down $\beta$'s sign in
Theorem~\ref{thm:uniqueness}. \textbf{Case 2 ($k$ affects some, not all,
groups).} Let $\delta\ne0$ be the shared change, $g^*$ unaffected. Factoring
$e^{\beta\delta}$ out of affected terms:
$$
w'_{i,g}=\frac{e^{\beta c_{i,g}}}{\sum_{\text{aff.}}e^{\beta c_{i,g'}}+
e^{-\beta\delta}\sum_{\text{unaff.}}e^{\beta c_{i,g''}}}.
$$
$\delta>0\Rightarrow e^{-\beta\delta}<1\Rightarrow w'_{i,g}>w_{i,g}$;
$\delta<0$ gives the mirror inequality. \qed
\end{proof}

\begin{remark}[Necessity of the proviso]
Without it: $\mathcal{G}_i=\{g,h\}$, $M_g\setminus\{i\}=M_h\setminus\{i\}=
\{k\}$ gives $c_{i,g}=c_{i,h}$ identically, $w_{i,g}=w_{i,h}=\tfrac12$
regardless of $\phi_{ik}$ --- not a defect, since a signal moving every
group identically carries no relative information.
\end{remark}

\begin{requirement}[Locality]
\label{req:locality}
If two parts of the player set have zero mutual interaction, weights in one part are unaffected by anything in the other: $N=N_1\sqcup N_2$, $\phi_{jk}=0$ for $j\in N_1,k\in N_2$: for $i\in N_1$, $w_{i,g}$ is unaffected by any $\phi_{kl}$, $k,l\in N_2$.
\end{requirement}

This asks for something close to common sense: if two groups of players
have absolutely nothing to do with each other --- every cross-term between
them is zero --- then changing interactions that happen entirely on the
other side should not somehow reach across and change anyone's weights on
this side. A rule that violated this would be reacting to information it
has no business reacting to.

\begin{proof}
$c_{i,g}=\sum_{k\in N_1\setminus\{i\}}\phi_{ik}+\underbrace{\sum_{k\in N_2}\phi_{ik}}_{=0}$, so $c_{i,g}$ depends only on $N_1$ internal terms. \qed
\end{proof}

\begin{requirement}[Continuity]
\label{req:continuity}
A tiny change in any interaction produces only a
tiny change in any weight: $\{\phi_{ij}\}\mapsto\{w_{i,g}\}$ is continuous everywhere.
\end{requirement}

This rules out a rule that jumps discontinuously --- say, flipping a player's entire weight from one group to another after an infinitesimally small change in one interaction term. Interaction data in practice is rarely known with perfect precision, so a rule that is wildly sensitive to tiny measurement noise would be unusable.

\begin{proof}
$c_{i,g}$ is a finite sum of linear terms, hence continuous. Formula~\eqref{eq:softmax} composes the exponential with a sum and division, continuous since the
denominator is strictly positive (Proposition~\ref{prop:validity}). \qed
\end{proof}

\begin{requirement}[Universal Validity]
\label{req:validity}
Weights must always be valid, for every interaction profile and every sign pattern --- the property (NAIVE), and
every rule in family~\eqref{eq:ratiorule}, provably lack: For every $i$ and nonempty $S\subseteq\mathcal{G}_i$: $w_{i,g}^S>0$ for $g\in S$, $\sum_{g\in S}w_{i,g}^S=1$. Stated for a general candidate rule, prior to any softmax assumption, since Section~\ref{sec:uniqueness} needs it as a positivity condition before formula~\eqref{eq:softmax} is known to be the answer.
\end{requirement}

This is simply Section~\ref{sec:construction}'s (M1)--(M2) validity demand,
restated here as a requirement so that Section~\ref{sec:uniqueness} can
build on it directly, using the sub-menu notation $w_{i,g}^S$ already set
up in Definition~\ref{def:fuzzy-membership} --- notation that becomes
essential once Requirement~\ref{req:iia} is introduced.

\begin{proposition}[Sensitivity to global sign-flip]
\label{prop:signflip}
Under $\phi_{ij}\to-\phi_{ij}$ for all $i\neq j$, $w_{i,g}$ changes for any
$i$ with non-constant $\{c_{i,g}\}_{g\in\mathcal{G}_i}$.
\end{proposition}

This last fact is a sanity check rather than a formal requirement: if every
interaction in the game were flipped in sign --- every synergy became a
redundancy and vice versa --- a player's weights should generally change
too, rather than staying suspiciously fixed. A rule that ignored a
wholesale sign-flip of its only input would clearly not be using that input
in any meaningful way.

\begin{proof}
$c_{i,g}\to-c_{i,g}$; weights coincide iff $e^{2\beta c_{i,g}}$ is constant
across $\mathcal{G}_i$ --- a degenerate special case, not the typical one.
\qed
\end{proof}
%%%%%%%%%%%%%%%%%%%%%%%%%%%%%%%%%%%%%%%%%%%%%%%%%%%%%%%%%%%%%%%%%%%%%%%%
\section{Uniqueness}
\label{sec:uniqueness}

Section~\ref{sec:axioms} gave six properties and checked that
formula~\eqref{eq:softmax} has all of them. That only shows softmax is
\emph{one valid answer}; it does not rule out some entirely different
formula also having all six properties. This section closes that gap: it
adds four more requirements and shows that, together, the full list leaves
no room for any rule other than softmax --- not merely \emph{a} rule with
the right properties, but, under a condition defined below, \emph{the
only} one.

\textbf{Proof sketch.} The argument runs in five stages, each one
narrowing down what a valid rule could possibly look like until only
softmax is left standing. (1) Universal Validity and Independence of
Irrelevant Groups, via a self-contained proof of Luce's representation,
force a positive per-label scale consistent across sub-menus --- in plain
terms, every group effectively gets assigned a single positive ``weight
unit,'' and comparing two groups never depends on which other groups
happen to be in the running. (2) Value Sufficiency forces this scale to be
a single function of the underlying real signal --- the weight unit for a
group can depend only on that group's own raw number, nothing else. (3) Decomposability, together with a precisely defined richness condition, forces this function to satisfy the multiplicative Cauchy equation --- disjoint, separable pieces of supporting evidence must combine multiplicatively, and continuity forces that composition rule into one specific shape. (4) Continuity forces its solutions to be exponential, for some $\beta\in\mathbb{R}$ --- ruling out the handful of pathological, nowhere-continuous solutions the Cauchy equation technically allows. (5) Monotonicity (Requirement~\ref{req:mono}) rules out $\beta\le0$, recovering
formula~\eqref{eq:softmax} exactly --- the last remaining ambiguity, the \emph{sign} of the exponential's rate, is pinned down by requiring weights to actually respond the right way to stronger synergy.

\subsection{Independence of Irrelevant Groups}
\label{sec:iia}

The first new requirement borrows an idea from choice theory: how a player
splits its attention between two particular groups should not depend on
which \emph{other} groups happen to also be under consideration at the
same time.

\begin{requirement}[Independence of Irrelevant Groups]
\label{req:iia}
Whether $g$ and $h$ are compared within the full set of $i$'s known groups or within any smaller sub-menu that still contains both, the relative weight $i$ assigns between them should not change: for any nonempty $S\subseteq\mathcal{G}_i$, $g,h\in S$:
$w_{i,g}^S/w_{i,h}^S=w_{i,g}^{\mathcal{G}_i}/w_{i,h}^{\mathcal{G}_i}$.
\end{requirement}

Concretely: if we ask how a legislator splits attention just between
Finance and Budget, ignoring Rules entirely, that two-way split should
come out the same whether or not Rules is also part of the picture. A rule
that let irrelevant alternatives shift the balance between two options
would be reacting to something it has no business reacting to. This exact
property, applied to probabilistic choice, is Luce's Choice Axiom; the
proposition below adapts it to our deterministic setting and shows it
forces every group's weight to come from one shared, per-group ``scale''
number:

\begin{proposition}[Luce's representation, finite deterministic case]
\label{prop:luce}
Suppose $\{p_S\}_{\emptyset\ne S\subseteq\mathcal{G}_i}$ satisfies (i)
$\sum_{g\in S}p_S(g)=1$ (Requirement~\ref{req:validity}) and (ii)
$p_S(g)/p_S(h)=p_{\mathcal{G}_i}(g)/p_{\mathcal{G}_i}(h)$
(Requirement~\ref{req:iia}). Setting $u(g):=p_{\mathcal{G}_i}(g)$,
$p_S(g)=u(g)/\sum_{h\in S}u(h)$ for every nonempty $S$ and $g\in S$.
\end{proposition}

\begin{proof}
Fix $S,g\in S$. By (ii), $p_S(h)=p_S(g)u(h)/u(g)$ for each $h\in S$.
Summing over $h\in S$ and using (i) for this $S$: $1=(p_S(g)/u(g))
\sum_{h\in S}u(h)$, giving $p_S(g)=u(g)/\sum_{h\in S}u(h)$. \qed
\end{proof}

This is a complete, self-contained proof: it uses only that $p_S(\cdot)$ is
a positive, unit-sum vector for each menu, an exact deterministic identity,
not Luce's original stochastic-choice argument. What it buys us is
significant: every group $g$ can be assigned one number $u(g)$, and every
sub-menu's weights are just that group's number divided by the sum of
numbers over the sub-menu. The next question is what $u(g)$ is actually
allowed to depend on.

\subsection{Value Sufficiency and Portability}
\label{sec:value-suff}

Proposition~\ref{prop:luce} hands us a per-group number $u(g)$, but it says
nothing about where that number comes from. In principle $u(g)$ could
depend on the group's label, on which other groups happen to exist, on the
particular player --- almost anything. The next requirement pins this
down: a group's scale should depend on nothing but that one group's own
raw signal $c_{i,g}$, through a single function that is the same across
every group and every situation for a given player.

\begin{requirement}[Value Sufficiency and Portability]
\label{req:value-suff}
The scale granted to a group should depend on nothing but that group's own raw signal, through one function fixed for the player and reusable across every profile: there is a single $u_i:\mathbb{R}\to(0,\infty)$, unique up to one positive scalar, such that the scale granted to any label $g\in\mathcal{G}_i$ by Requirement~\ref{req:iia}, under any admissible profile, equals $u_i(c_{i,g})$.
\end{requirement}

Without this requirement, nothing would stop the rule from treating
Finance and Rules differently purely because of their \emph{names}, even
when the underlying numbers --- how synergistic or redundant a player is
with each committee --- are identical. The remark and witness below show
this is not an automatic consequence of the two requirements proved so
far; it has to be asked for separately.

\begin{remark}
With two labels, a weaker cross-profile ratio-matching condition reduces to
$R(x,y)R(y,x)=1$, satisfied by $R(x,y)=e^{\sin(x-y)}$, yet no single $u$
gives this. We posit Requirement~\ref{req:value-suff} directly.
\end{remark}

\textit{Witness.} Fix any total order on the universe of groups $G$,
independent of player or profile, and let $\mathrm{rank}(g)$ be $g$'s
position in that order. For fixed $\varepsilon\in(0,1)$ define
$\kappa(g):=1+\varepsilon\cdot\mathrm{rank}(g)$ and, for every player and
every sub-menu $S\subseteq\mathcal{G}_i$,
$$w_{i,g}^S:=\frac{e^{\beta c_{i,g}}\kappa(g)}{\sum_{g'\in S}e^{\beta
c_{i,g'}}\kappa(g')}.$$
This rule satisfies Requirement~\ref{req:validity} ($\kappa(g)>0$, so the
argument of Proposition~\ref{prop:validity} applies verbatim),
Requirement~\ref{req:iia} (the family is already indexed by $S$ in the
required ratio-consistent form), Requirement~\ref{req:continuity}
($\kappa$ does not depend on $\phi$, so continuity in $c_{i,g}$ transfers
unchanged), Requirement~\ref{req:scale} ($\kappa(g)$ is unaffected by
$\phi\to\lambda\phi$, so the same cancellation as in
Requirement~\ref{req:scale}'s proof applies), Requirement~\ref{req:locality}
($\kappa(g)$ never references any $\phi_{kl}$ term), and
Requirement~\ref{req:mono} ($\partial w_{i,g}^S/\partial
c_{i,g}=\beta w_{i,g}^S(1-w_{i,g}^S)$ exactly as in
Requirement~\ref{req:mono}'s own proof, since $\kappa(g)$ is constant with
respect to $c_{i,g}$). Yet it violates Requirement~\ref{req:value-suff}:
two groups $g,k$ with identical raw signal $c_{i,g}=c_{i,k}$ but different
ranks get different scales, $e^{\beta c_{i,g}}\kappa(g)\ne e^{\beta
c_{i,g}}\kappa(k)$, so no single function of $c_{i,g}$ alone could have
produced these weights. Concretely, with $\mathcal{G}_i=\{g,h,k\}$,
$c_{i,g}=1,c_{i,h}=2,c_{i,k}=1$, and $\mathrm{rank}(g)\ne\mathrm{rank}(k)$,
this rule gives $w_{i,g}\neq w_{i,k}$ despite $c_{i,g}=c_{i,k}$: two groups
with the \emph{identical} raw signal end up with different weights, exactly
the arbitrary label-dependence Value Sufficiency is designed to rule out.

\subsection{Richness}

The remaining two requirements --- Decomposability, below, and the
uniqueness theorem that follows it --- need one more piece of structure
about the group layout itself, not about the rule. Loosely: to pin down exactly how a group's scale function behaves, we need at least one group with enough co-members that we can split its supporting evidence into two disjoint, separable pieces and test how the rule combines them.

\begin{definition}[Rich player]
\label{def:rich}
Player $i$ is \emph{rich} if there is a \emph{witnessing group}
$g_0\in\mathcal{G}_i$ with $|M_{g_0}\setminus\{i\}|\ge2$ and at least one
\emph{exclusive witness} $k\in M_{g_0}\setminus\{i\}$ with
$\mathcal{G}_i\cap\mathcal{G}_k=\{g_0\}$ --- $k$ shares \emph{only} $g_0$
with $i$.
\end{definition}

In words: $i$ is rich if at least one of its groups has at least two
co-members in total, and at least one of them has \emph{no other group in
common with $i$}. The size condition is what lets the group's evidence be split into two disjoint, separable pieces (Lemma~\ref{lem:realizability-cauchy} below); the single exclusive co-member is what lets us vary $i$'s signal toward $g_0$ without disturbing $i$'s signal toward any other group
(Lemma~\ref{lem:realizability-continuity} below). Only one exclusive co-member is needed for this: any further co-members of $g_0$, exclusive or not, play no role in the argument.

\begin{definition}[Eligible player]
\label{def:eligible}
Player $i$ is \emph{eligible} if $i$ is rich and $|\mathcal{G}_i|\ge2$.
\end{definition}

Richness alone is not quite enough: the theorem below also needs a player
to belong to \emph{at least two} groups, since the whole point of
``attribution'' is comparing a group against some other group, and a
player with only one group has nothing to compare against.

Neither clause implies the other: a player belonging to a single group
with two co-members is rich but not eligible (Remark~\ref{rem:degeneracy});
Remark~\ref{rem:richness} exhibits $|\mathcal{G}_i|=2$ without richness.
Exclusivity of the witnesses is what lets the realizability argument below
vary a group's evidence without perturbing any other group's --- a bare
co-membership count does not.

\subsection{Decomposability}

With richness defined, we can state the requirement that does the heaviest mathematical lifting in this paper: how a group's scale number should behave when its supporting evidence can be split into two disjoint, separable pieces.

\begin{requirement}[Decomposability of Separable Evidence]
\label{req:decomp}
\textbf{Setup.} By Requirements~\ref{req:iia}--\ref{req:value-suff},
$w_{i,g}=u_i(c_{i,g})/\sum_{g'}u_i(c_{i,g'})$, unique up to a scalar
(Lemma~\ref{lem:normalization}); normalize this scalar ambiguity by fixing
$u_i(0)=1$.

When a group's supporting evidence splits into two disjoint, separable parts, combining them should multiply their separate scales rather than require joint re-evaluation: for any nonempty $A,B$ with $M_g\setminus\{i\}=A\sqcup B$: $u_i(c_{i,g}^A+c_{i,g}^B)=u_i(c_{i,g}^A)\cdot u_i(c_{i,g}^B)$.
\end{requirement}

To be precise about what this requirement does and does not claim: it
does \emph{not} assert that $c_{i,g}$ is a likelihood, or that cooperative-game
interactions carry probabilistic semantics. What it claims is narrower and
purely structural: $c_{i,g}$ is already an \emph{additive} signal
(equation~\eqref{eq:rawsignal} sums pairwise terms), so when a group's
co-members split into two disjoint subsets, each contributing its own
piece of that sum, we ask that the \emph{scale} assigned to the combined
signal compose consistently with the scales assigned to its two separable
pieces --- multiplicatively, since the pieces combine additively at the
signal level and the scale must stay positive throughout. The same
algebraic pattern appears in how independent likelihoods multiply in
Bayesian updating, which motivates the choice of composition rule but is
not the semantics being imposed here. This requirement, together with richness, is what forces the exponential shape; the next three results carry out that argument step by step. Decomposability is a substantive modeling choice, not a consequence of cooperative-game theory alone: Independence of Irrelevant Groups and Value Sufficiency already force the ratio form $w_{i,g}\propto u_i(c_{i,g})$
(Section~\ref{sec:iia}--\ref{sec:value-suff}) for \emph{some} positive function $u_i$; Decomposability is what additionally selects the exponential member of that broader class. A positive, continuous, strictly increasing function can satisfy every other requirement while violating Decomposability --- for instance $u(x)=e^x+1$ is positive, continuous, and strictly increasing, yet $u(0+0)=2\ne4=u(0)u(0)$ --- so
this requirement is doing genuine, non-vacuous work, not restating a consequence of the others.

The proof of Theorem~\ref{thm:uniqueness} uses richness in two logically separate ways, so we isolate them as two lemmas rather than one. Decomposability (Lemma~\ref{lem:cauchy} below) only needs $g_0$ to have
\emph{some} nonempty two-part split of its co-members, and makes no use of
exclusivity at all:

\begin{lemma}[Realizability for Decomposability]
\label{lem:realizability-cauchy}
Let $g_0\in\mathcal{G}_i$ have $|M_{g_0}\setminus\{i\}|\ge2$, and fix any
nonempty partition $M_{g_0}\setminus\{i\}=A\sqcup B$. For any real $x,y$,
setting the $\phi_{i\ell}$ terms for $\ell\in A$ to sum to $x$ and those
for $\ell\in B$ to sum to $y$ (other terms at an arbitrary baseline)
realizes any target pair $(x,y)$.
\end{lemma}

\begin{proof}
$c_{i,g_0}=\sum_{\ell\in A}\phi_{i\ell}+\sum_{\ell\in B}\phi_{i\ell}$; each
sum is a free real number once the other co-members' terms are fixed at
baseline, so $x,y$ range freely over $\mathbb{R}^2$. \qed
\end{proof}

The continuity transfer in Step 3 of Theorem~\ref{thm:uniqueness}'s proof
needs something different: a way to move $c_{i,g_0}$ over all of
$\mathbb{R}$ \emph{without disturbing} $c_{i,h}$ for any other group
$h$. This is exactly what a single exclusive witness provides:

\begin{lemma}[Realizability for Continuity]
\label{lem:realizability-continuity}
Let $i$ be rich with witnessing group $g_0$ and exclusive witness $k$. For
any real $x$, setting $\phi_{i,k}=x$ (other terms at an arbitrary
baseline) realizes any target value $x$ for $c_{i,g_0}$ while leaving
$c_{i,h}$ fixed for every $h\in\mathcal{G}_i\setminus\{g_0\}$.
\end{lemma}

\begin{proof}
$c_{i,g_0}=\phi_{i,k}+\sum_{\ell\in M_{g_0}\setminus\{i,k\}}\phi_{i\ell}$;
the sum is fixed at baseline, so $x$ ranges freely. For $h\ne g_0$:
exclusivity gives $k\notin M_h$, so $c_{i,h}$ contains no $\phi_{ik}$ term
and is unaffected. \qed
\end{proof}

Together these two lemmas are what makes richness useful: the size condition on $g_0$ lets its evidence be split into two disjoint, separable pieces for Decomposability, and the single exclusive witness lets that group's signal be varied in isolation for continuity --- two distinct jobs, each requiring only what it is given here.

\begin{lemma}[Requirement~\ref{req:decomp} forces the Cauchy equation]
\label{lem:cauchy}
Given richness, $u_i(x+y)=u_i(x)u_i(y)$ for all real $x,y$.
\end{lemma}

\begin{proof}
Richness guarantees $|M_{g_0}\setminus\{i\}|\ge2$; fix any nonempty
partition $A\sqcup B$ of $M_{g_0}\setminus\{i\}$. By
Lemma~\ref{lem:realizability-cauchy}, Requirement~\ref{req:decomp} gives
$u_i(x+y)=u_i(x)u_i(y)$ for all real $x,y$ directly. \qed
\end{proof}

The equation $u_i(x+y)=u_i(x)u_i(y)$ is the multiplicative Cauchy
equation, one of the oldest functional equations in mathematics, and ---
once we add continuity, in the theorem below --- its only solutions are
exponentials. One small bookkeeping step remains before we can invoke
that fact: fixing exactly which exponential, by pinning down $u_i(0)$.

\begin{lemma}[Normalization of $u_i$]
\label{lem:normalization}
$\tilde u_i:=u_i/u_i(0)$ satisfies $\tilde u_i(0)=1$, using only
$u_i(0)>0$ (Requirement~\ref{req:value-suff}'s codomain $(0,\infty)$) ---
no appeal to Requirement~\ref{req:decomp}.
\end{lemma}

\subsection{The Uniqueness Theorem}

Everything above now assembles into the paper's central result: for an
eligible player, the five requirements proved so far leave exactly one
possible shape for $u_i$ --- an exponential --- and the sixth,
Monotonicity, pins down the sign of its rate.

\begin{theorem}[Uniqueness of restricted-domain softmax]
\label{thm:uniqueness}
Fix an eligible player $i$ (Def.~\ref{def:eligible}). Among all
menu-indexed families $\{c_{i,\cdot}\mapsto w_{i,\cdot}^S\}_{\emptyset\ne
S\subseteq\mathcal{G}_i}$ (Def.~\ref{def:fuzzy-membership}) satisfying
Requirements~\ref{req:continuity}, \ref{req:validity}, \ref{req:iia},
\ref{req:value-suff}, and \ref{req:decomp}, there is a single
$\beta\in\mathbb{R}$ with $u_i(x)=e^{\beta x}$ and, on every such $S$,
$w_{i,g}^S$ given by \eqref{eq:softmax}'s functional form. If
Requirement~\ref{req:mono} is additionally required, $\beta$ is forced
strictly positive, recovering formula \eqref{eq:softmax} exactly on every
sub-menu, including the full menu $S=\mathcal{G}_i$.
\end{theorem}

\begin{proof}
\textbf{Steps 1--2 (uses only richness).} By Prop.~\ref{prop:luce} and
Requirement~\ref{req:value-suff}, $u(g)=u_i(c_{i,g})$, one function; by
Lemma~\ref{lem:cauchy}, $u_i(x+y)=u_i(x)u_i(y)$ on all reals.

\textbf{Step 3 (continuity transfer --- needs $|\mathcal{G}_i|\ge2$).}
Requirement~\ref{req:continuity} gives continuity of the full vector map,
not directly of the scalar $u_i$. Fix a baseline and let
$K:=\sum_{h'\in\mathcal{G}_i\setminus\{g_0\}}u_i(c_{i,h'}^{\text{baseline}})$
--- a finite sum of positive terms, hence $K>0$, nonempty since $i$ is
eligible. By Lemma~\ref{lem:realizability-continuity}, varying
$\phi_{i,k}$ moves
$x:=c_{i,g_0}$ over $\mathbb{R}$ while holding every term contributing to
$K$ fixed, so $G(x):=w_{i,g_0}=u_i(x)/(u_i(x)+K)$ is continuous in $x$.
Inverting the continuous bijection $t\mapsto t/(t+K)$ gives
$u_i(x)=K\cdot G(x)/(1-G(x))$, continuous.

\textbf{Step 4 (Cauchy + continuity).} $u_i(0)=1$
(Lemma~\ref{lem:normalization}); $u_i(-x)=1/u_i(x)$; $u_i(n)=u_i(1)^n$ on
integers, rationals; continuity (Step 3) extends this to $u_i(x)=u_i(1)^x$
on all of $\mathbb{R}$. Writing $u_i(1)=e^\beta$: $u_i(x)=e^{\beta x}$.

\textbf{Step 5 (substitution).} $w_{i,g}=e^{\beta c_{i,g}}/\sum_{g'}
e^{\beta c_{i,g'}}$, valid for any $\beta\in\mathbb{R}$ so far.

\textbf{Step 6 (Requirement~\ref{req:mono} fixes the sign --- needs
$|\mathcal{G}_i|\ge2$).} Take $g^*\ne g_0$, which exists since $i$ is
eligible; $k\notin M_{g^*}\setminus\{i\}$ by exclusivity, satisfying
Requirement~\ref{req:mono}'s hypothesis. If $\beta\le0$:
$\partial w_{i,g_0}/\partial c_{i,g_0}=\beta w_{i,g_0}(1-w_{i,g_0})\le0$,
contradicting strict increase. Hence $\beta>0$. \qed
\end{proof}

A quick summary of the chain just completed: Luce's representation (Steps
1--2) gets us a single scale function $u_i$; the Cauchy equation (built
from Decomposability and richness) plus continuity (Steps 3--4) force that
function to be exponential; and only the \emph{sign} of the exponent's
rate is left open after Step 5, which Monotonicity (Step 6) settles. The
two remarks below show both halves of this argument are load-bearing ---
dropping either richness or Monotonicity genuinely breaks the result,
rather than being harmless simplifications.

\begin{remark}
\label{rem:mono-independence}
Requirements~\ref{req:continuity}--\ref{req:decomp}, together with
eligibility, force the exponential family, not its sign: $\beta=0$
(uniform, ignoring the interaction signal) and $\beta<0$ (concentrating on
the \emph{least} synergistic group) both satisfy them. Only
Requirement~\ref{req:mono} excludes these, which simultaneously shows
Requirement~\ref{req:mono}'s independence from the rest: the $\beta\le0$
rules satisfy every other requirement for an eligible player while
violating this one.
\end{remark}

\begin{remark}[Richness is not vacuous]
\label{rem:richness}
$\mathcal{G}_1=\{g,h\}$, $M_g\setminus\{1\}=\{2\}$, $M_h\setminus\{1\}=
\{3\}$: $|\mathcal{G}_1|=2$ but richness fails (no group has two
co-members). $u_1(x)=2+\arctan(x)$ satisfies every requirement including
Monotonicity, yet at $(1,0)$ gives $w_{1,g}\approx0.5822$ versus softmax's
$\approx0.7311$ --- genuinely non-exponential.
\end{remark}

\begin{remark}[The $|\mathcal{G}_i|=1$ degeneracy]
\label{rem:degeneracy}
A player rich via a single group ($\mathcal{G}_i=\{g_0\}$,
$M_{g_0}\setminus\{i\}=\{k_1,k_2\}$) is not eligible: $K$ becomes an empty
sum ($=0$), and formula~\eqref{eq:softmax} itself gives $w_{i,g_0}\equiv1$ for \emph{every} $\beta$, since (M2) forces the single weight to 1 before $\beta$ enters. Requirement~\ref{req:mono}'s hypothesis is unsatisfiable
here too (no comparison group exists), so it imposes no constraint. This is
why eligibility's second clause, not richness alone, is necessary.
\end{remark}

A practical by-product of the theorem is worth stating on its own: once a
rule is known to be softmax of \emph{some} temperature, that temperature
can be read directly off any two groups with different signals, no
re-derivation needed.

\begin{lemma}[Recoverability of $\beta$]
\label{lem:recoverability}
If $c_{i,g}\neq c_{i,h}$: $\beta=\ln(w_{i,g}/w_{i,h})/(c_{i,g}-c_{i,h})$.
If $c_{i,\cdot}$ is constant, every $\beta$ gives the uniform weight
vector on this profile.
\end{lemma}

\begin{proof}
$w_{i,g}/w_{i,h}=e^{\beta(c_{i,g}-c_{i,h})}$; logs recover $\beta$
uniquely when $c_{i,g}\ne c_{i,h}$. Constant $c_{i,\cdot}=c_0$: every term
$e^{\beta c_0}$ is identical regardless of $\beta$. \qed
\end{proof}

\subsection{A Single Shared Temperature Across Players}

Everything so far pins down a temperature $\beta_i$ \emph{per player} ---
nothing yet stops two different players from ending up with two different
rates of sensitivity to their own groups' signals. The final requirement
closes that last gap by asking for one shared $\beta$ across the whole
population.

\begin{example}[Independence witness for Rule Universality]
\label{ex:group-count-beta}
$\beta_i:=1+|\mathcal{G}_i|$ --- exogenous taxonomy data, never depending
on $\phi$; $w_{i,g}:=e^{\beta_i c_{i,g}}/\sum_{g'}e^{\beta_i c_{i,g'}}$.
\end{example}

This example shows the gap is real: a rule can satisfy every single
requirement proved so far while still letting $\beta$ silently vary from
player to player, as long as the variation is tied to something harmless
like how many groups a player belongs to.

\begin{remark}
An earlier candidate witness, $\beta_i:=1/\bar c_i$ with $\bar c_i$ the
mean absolute signal, is computed \emph{from} the profile itself and
therefore violates Requirement~\ref{req:value-suff}: it is not an
isolating witness. The construction above depends only on $|\mathcal{G}_i|$,
fixed for a given player across every profile, and is verified against
every other requirement directly: it inherits Requirements
\ref{req:continuity}--\ref{req:decomp} by being literally
$\mathrm{softmax}_{\beta_i}$ for a positive constant; Symmetry, since
symmetric twins share $|\mathcal{G}_i|$; Scale Invariance, since $\beta_i$
does not depend on $\phi$'s scale; Locality, since $\beta_i$ never
references any $\phi_{kl}$ term; and Monotonicity, since $\beta_i\ge2>0$
always. It violates Rule Universality: players with differing
$|\mathcal{G}_i|$ receive differing $\beta_i$, with nothing forcing them equal.
\end{remark}

A second independence witness is worth including here, since it isolates
Decomposability specifically (the requirement doing the most mathematical
work in this section) using the paper's own flagship example rather than
an abstract construction.

\begin{remark}[Independence witness for Requirement~\ref{req:decomp}]
\label{rem:decomp-independence}
Fix any eligible player $i$ with witnessing group $g_0$ and exclusive
witnesses $k_1,k_2$ (in the flagship instance of
Section~\ref{sec:empirical}: player 3, via Rules $g_3$, witnessed by
players 5 and 6). Let $u(x):=e^x(1+\varepsilon x^2)$ for any
$0<\varepsilon<1$ (e.g.\ $\varepsilon=0.1$), and define
$w_{i,g}:=u(c_{i,g})/\sum_{g'\in\mathcal{G}_i}u(c_{i,g'})$ for every
player and every menu.

This rule satisfies Requirement~\ref{req:validity} ($u(x)>0$ for all real
$x$, so the argument of Proposition~\ref{prop:validity} applies
verbatim), Requirement~\ref{req:iia} (by construction
$w_{i,g}/w_{i,h}=u(c_{i,g})/u(c_{i,h})$ for any sub-menu, never
referencing the menu itself), Requirement~\ref{req:value-suff} (a single
function $u$, identical across every player and profile), and
Requirement~\ref{req:continuity} ($u$ is smooth and the denominator is a
finite sum of strictly positive terms, hence bounded away from $0$).

Yet it violates Requirement~\ref{req:decomp}. Using player 3's own
witnesses and the flagship instance's own edge weights, $\phi_{3,5}=
-0.30$, $\phi_{3,6}=0.15$:
\begin{center}
\begin{tabular}{@{}cccc@{}}
\toprule
$x$ & $y$ & $u(x+y)$ & $u(x)\cdot u(y)$ \\
\midrule
$-0.30$ & $0.15$ & $0.86264$ & $0.87041$ \\
$0.40$ & $-0.60$ & $0.82201$ & $0.86178$ \\
$1.00$ & $-1.00$ & $1.00000$ & $1.21000$ \\
\bottomrule
\end{tabular}
\end{center}
The gap is not an artifact of one chosen pair --- it holds across the
profile's own realizable values --- so this is a genuine, non-vacuous
witness: Requirement~\ref{req:decomp} is independent of
Requirements~\ref{req:continuity}, \ref{req:validity},
\ref{req:iia}, and \ref{req:value-suff}. Unlike Remark~\ref{rem:richness}'s
$\arctan$ example, this witness is built on a \emph{rich} player, so
Lemma~\ref{lem:realizability-cauchy}'s two-part split genuinely applies and
Requirement~\ref{req:decomp} is exercised rather than vacuously satisfied.
\end{remark}

With the gap demonstrated, the final requirement closes it directly:

\begin{requirement}[Rule Universality]
\label{req:universality}
Nothing about a player's individual identity, only the shared population
of players, should determine how sharply group-signal differences convert
into weight: restricted to eligible players, the $\{\beta_i\}$ granted by
Theorem~\ref{thm:uniqueness} satisfy $\beta_i=\beta_j$ for all $i,j$.
\end{requirement}

Putting this together with everything proved so far gives the paper's
final form: one rule, one shared temperature, no per-player subscripts
left anywhere.

\begin{corollary}[Formula~\eqref{eq:softmax} with one shared $\beta$]
\label{cor:shared-beta}
If every player is eligible, Requirements~\ref{req:continuity}--\ref{req:decomp}
and \ref{req:mono} (each granting $\beta_i>0$) together with
Requirement~\ref{req:universality} give one $\beta\in\mathbb{R}_{>0}$ with
formula~\eqref{eq:softmax} exactly, no subscript.
\end{corollary}

\begin{proof}
Theorem~\ref{thm:uniqueness} per eligible player gives $\beta_i>0$;
Requirement~\ref{req:universality} identifies all $\beta_i$ as one common
value $\beta>0$. \qed
\end{proof}

This corollary requires \emph{every} player to be eligible. In the
flagship instance of Section~\ref{sec:empirical}, players 2 and 3 are
eligible (Section~\ref{subsec:graph-game}), but players 1, 4, 5, and 6
each have $|\mathcal{G}_i|=1$, failing eligibility outright. The corollary
therefore does not apply to that instance, and the shared $\beta=1$ used
in Tables~\ref{tab:softmax-out}--\ref{tab:degenerate} is a choice made for
the illustration, not a value forced by this corollary.

\subsection{Status of Each Requirement}

Having stated and used ten requirements across Sections~\ref{sec:axioms}
and~\ref{sec:uniqueness}, it is worth being explicit about how strong a
claim each one actually makes, since not all of them carry the same kind
of justification.

Requirements~\ref{req:iia} and \ref{req:decomp} are each motivated by an independent principle from another field: Luce's theorem for the former, and for the latter the same intuition given in Section~\ref{sec:uniqueness}'s discussion of Decomposability --- that disjoint, separable pieces of evidence should combine multiplicatively on the attribution scale, the way independent likelihoods multiply in Bayesian updating, rather than add. Requirements~\ref{req:continuity} and \ref{req:validity} are basic well-posedness conditions.
Requirement~\ref{req:mono} is a plain-language-compelling requirement in its own right, and is additionally shown independent of Requirements~\ref{req:continuity}--\ref{req:decomp} above. Requirements \ref{req:value-suff} and \ref{req:universality} are convention-like:
motivated by analogy rather than derived. Independence has been shown via explicit, fully verified counterexample for Requirement~\ref{req:mono} (Remark~\ref{rem:mono-independence}), Requirement~\ref{req:value-suff} (the Witness above), Requirement~\ref{req:decomp} (Remark~\ref{rem:decomp-independence}), and Requirement~\ref{req:universality} (Example~\ref{ex:group-count-beta}); whether Requirements~\ref{req:continuity}, \ref{req:validity}, or \ref{req:iia} individually admit similar witnesses remains open.
%%%%%%%%%%%%%%%%%%%%%%%%%%%%%%%%%%%%%%%%%%%%%%%%%%%%%%%%%%%%%%%%%%%%%%%%
\section{Worked Example}
\label{sec:empirical}
Sections~\ref{sec:axioms} and~\ref{sec:uniqueness} proved a representation
theorem: under a stated list of requirements, formula~\eqref{eq:softmax}
is forced. A proof like that establishes \emph{what} construction is
correct, but it is easy to lose track, across all the abstract machinery,
of what the construction actually \emph{does} to concrete numbers.
Real-world applications of membership attribution typically have no
ground-truth coalition values $v(S)$ against which an attribution could be
scored, so there is no predictive benchmark to run here; this section
instead uses a controlled, exact cooperative game to illustrate the
construction's behavior rather than to claim empirical validation. It is a
single, fully worked example --- every number below is computed directly
from a hand-specified cooperative game, with no estimation or
approximation anywhere --- whose purpose is to make the theorems in
Sections~\ref{sec:axioms} and~\ref{sec:uniqueness} checkable by hand and
easier to hold in mind.

\subsection{An Exact Graph-Restricted Game}
\label{subsec:graph-game}
To get concrete numbers we first need a concrete cooperative game, i.e.\ an
actual value $v(S)$ for every subset $S$ of players. Rather than making up
numbers directly, we build $v$ from a simple graph: fix a pairwise
interaction $\phi_{ij}$ for every pair of players, and let a coalition's
total value just be the sum of the pairwise interactions among its
members, as a graph game \citep{myerson1977}:
$v(S):=\sum_{\{i,j\}\subseteq S}\phi_{ij}$, $v(\emptyset)=v(\{i\})=0$. This
choice is convenient rather than restrictive: by the telescoping identity
of Lemma~\ref{lem:mobius}, every dividend of order $\ge3$ vanishes and the
order-2 dividends equal exactly the chosen $\phi_{ij}$ --- so the
$\phi_{ij}$ we write down \emph{are}, by construction, the pairwise
interaction terms the rest of the paper reasons about, with nothing left
to compute or approximate.

\textbf{Instance.} Six legislators, three overlapping committees: Finance
$g_1=\{1,2,3\}$, Budget $g_2=\{2,3,4\}$, Rules $g_3=\{3,5,6\}$.
Affiliations: $\mathcal{G}_1=\{g_1\}$, $\mathcal{G}_2=\{g_1,g_2\}$,
$\mathcal{G}_3=\{g_1,g_2,g_3\}$, $\mathcal{G}_4=\{g_2\}$, $\mathcal{G}_5=
\{g_3\}$, $\mathcal{G}_6=\{g_3\}$. Edge weights as in
Table~\ref{tab:edges}.

\begin{table}[h]
\centering
\begin{tabular}{@{}cc|cc|cc@{}}
\toprule
pair & $\phi$ & pair & $\phi$ & pair & $\phi$ \\
\midrule
1--2 & 0.30 & 2--3 & 0.50 & 3--5 & $-0.30$ \\
1--3 & 0.40 & 2--4 & $-0.60$ & 3--6 & 0.15 \\
1--4 & 0 & 2--5 & 0 & 4--5 & 0 \\
1--5 & 0 & 2--6 & 0 & 4--6 & 0 \\
1--6 & 0 & 3--4 & $-0.40$ & 5--6 & 0.35 \\
\bottomrule
\end{tabular}
\caption{Edge weights defining the exact graph game.}
\label{tab:edges}
\end{table}

Player 3 is the instance's most interesting case: it is positively
synergistic with players 1 and 2 (Finance and Budget), but actively
redundant with player 5 (Rules) --- exactly the mixed-sign situation
Proposition~\ref{prop:linear-ratio-fails} shows breaks naive normalization,
and the first thing this example checks.

\textbf{Eligibility check, on real data.} Before computing any weights, it
is worth checking which players Theorem~\ref{thm:uniqueness}'s uniqueness
guarantee actually covers, since not every player in this instance turns
out to be eligible. Player 3 is eligible: Rules ($g_3$) has
$|M_{g_3}\setminus\{3\}|=2\ge2$, and player 5 is an exclusive witness
($\mathcal{G}_5=\{g_3\}$ shares nothing else with player 3). Player 2 is
also eligible, once richness is stated at its minimal sufficient strength
(Definition~\ref{def:rich}): take $g_0=g_1$, with
$M_{g_1}\setminus\{2\}=\{1,3\}$, so $|M_{g_1}\setminus\{2\}|=2$; player 1
is an exclusive witness, since $\mathcal{G}_1=\{g_1\}$ shares nothing else
with player 2. Player 3 is \emph{not} a usable witness for $g_1$, since
$\mathcal{G}_2\cap\mathcal{G}_3=\{g_1,g_2\}\ne\{g_1\}$ --- but
Definition~\ref{def:rich} only requires \emph{one} exclusive witness per
witnessing group, and player 1 already supplies it.
Theorem~\ref{thm:uniqueness} therefore applies unconditionally to both
players 2 and 3 in this instance. Players 1, 4, 5, and 6 remain outside
the theorem's guarantee, since each has $|\mathcal{G}_i|=1$ and so fails
eligibility's second clause regardless of richness.

\subsection{Naive Normalization Fails}
Table~\ref{tab:naive-fail} applies rule~\eqref{eq:ratiorule} with
$h=\sum$ to players 2 and 3 of the flagship instance, both of whom mix
synergy and redundancy across their groups. If direct normalization worked,
every entry below would be a number between 0 and 1, reading as an ordinary
percentage.

\begin{table}[ht]
\centering
\begin{tabular}{@{}lccc@{}}
\toprule
Player & $w^{\text{NAIVE}}_{g_1}$ & $w^{\text{NAIVE}}_{g_2}$ & $w^{\text{NAIVE}}_{g_3}$ \\
\midrule
2 & 1.1429 & $-0.1429$ & --- \\
3 & 1.0588 & 0.1176 & $-0.1765$ \\
\bottomrule
\end{tabular}
\caption{Consistent with Prop.~\ref{prop:linear-ratio-fails}: both players
mix synergy and redundancy, so no linear-ratio rule can be valid here.}
\label{tab:naive-fail}
\end{table}

As Proposition~\ref{prop:linear-ratio-fails} predicted, that does not happen: player 3's Rules weight comes out to $-0.1765$, a negative percentage, and player 2's Finance weight exceeds $100\%$. These follow directly and inevitably from dividing a mixed-sign numerator by a shared denominator.

\subsection{The Construction's Output}
Table~\ref{tab:softmax-out} reports formula~\eqref{eq:softmax}'s output on
the same instance, at $\beta=1$, for both players. Every entry is now a
genuine percentage.

\begin{table}[ht]
\centering
\begin{tabular}{@{}lcccc@{}}
\toprule
Player & $\beta$ & $w_{g_1}$ & $w_{g_2}$ & $w_{g_3}$ \\
\midrule
2 & 1 & 0.7110 & 0.2890 & --- \\
3 & 1 & 0.5558 & 0.2497 & 0.1945 \\
\bottomrule
\end{tabular}
\caption{Valid throughout (Prop.~\ref{prop:validity}), regardless of
eligibility.}
\label{tab:softmax-out}
\end{table}

Notice what happened to player 3 in particular: despite being actively
redundant with Rules ($\phi_{35}=-0.30$), player 3 still receives a
positive, if small, weight there ($0.1945$) rather than a negative or
undefined one --- exactly the graceful handling of mixed-sign evidence the
softmax construction was built to provide.

\subsection{Numerical Verification of the Requirements}
With both tables in hand, we can now check each requirement from
Section~\ref{sec:axioms} directly against these numbers, one at a time.

\textbf{Universal Validity (Req.~\ref{req:validity}).} Confirmed above:
every row sums to $1.0000$, every entry in $(0,1)$.

\textbf{Scale Invariance (Req.~\ref{req:scale}).} Rescaling all $\phi_{ij}$
by $\lambda=2$, $\beta\to\beta/2$, exactly reproduces
Table~\ref{tab:softmax-out}'s row for player 2 --- doubling every
interaction and halving the temperature leaves the weights completely
unchanged, exactly as Requirement~\ref{req:scale} demands.

\textbf{Monotonicity (Req.~\ref{req:mono}).} Perturbing $\phi_{35}$ from
$-0.30$ to $-0.10$ moves $c_{3,g_3}$ from $-0.15$ to $0.05$; $w_{3,g_3}$
rises from $0.1945$ to $0.2278$, using $g^*=g_1$ (player $5$ is exclusive
to player 3's $g_3$, so $g_1$ is genuinely unaffected) as the required
comparison group. In words: as player 3 becomes less redundant with
(closer to neutral toward) Rules, Rules' share of player 3's weight
strictly increases, exactly as the requirement demands.

\textbf{Sign-flip sensitivity (Prop.~\ref{prop:signflip}).} Flipping every
sign of player 3's $\phi$-values gives $(0.1644,0.3658,0.4698)$ --- the
ranking inverts. Player 3's weight shifts from being concentrated on
Finance to being concentrated on Rules, confirming the construction
genuinely reacts to its input rather than ignoring it.

\textbf{Symmetry, Locality.} These two requirements are proved directly from the construction's algebraic form in Section~\ref{sec:axioms} rather than checked against a numeric instance there; no separate witness instance for either exists elsewhere in this paper. Symmetry does not apply within the flagship instance itself: although players 5 and 6 share identical affiliations ($\mathcal{G}_5=\mathcal{G}_6=\{g_3\}$), they differ in their interaction with player 3 ($\phi_{53}=-0.30\neq\phi_{63}=0.15$), so Requirement~\ref{req:symmetry}'s hypothesis is not met by any pair here. Locality is likewise untestable on this instance, since the nonzero edges connect all six players through player 3 and no zero-interaction partition $N_1\sqcup N_2$ exists. These two requirements simply do not have a witness pair or a disconnected component available within this particular six-player instance --- a limitation of the example, not of the requirements themselves, which is why Sections~\ref{sec:axioms} verifies them on separate, purpose-built instances instead.

\subsection{The Eligibility Boundary, Illustrated Twice}
The eligibility check earlier in this section showed that both players 2 and 3 are covered by Theorem~\ref{thm:uniqueness}'s uniqueness guarantee in this instance. This final subsection shows what the boundary looks like when it \emph{is} crossed --- first on a hypothetical addition to the flagship instance, then on a second, hand-picked example that pushes a different failure mode further.

\textbf{Coarse eligibility failure (a hypothetical player).} A player can have $\ge2$ co-members in a group and still fail richness, if \emph{every} one of them overlaps with the player's other groups. For instance, a hypothetical player 7 with $\mathcal{G}_7=\{g_1,g_4\}$ added to this instance, where $g_4=\{2,3,7\}$: both co-members of player 7 in $g_1$
(players 2 and 3) also sit in $g_4$, so neither is an exclusive witness and player 7 is not rich via $g_1$; symmetric reasoning rules out $g_4$ as a witnessing group too. Formula~\eqref{eq:softmax} would still compute a
valid weight for such a player (Prop.~\ref{prop:validity} needs no eligibility), but Theorem~\ref{thm:uniqueness} would not certify uniqueness for them --- some other, non-exponential rule could in principle also have satisfied every requirement for that player specifically.

\textbf{Fine eligibility failure (identical co-membership).} A sharper degeneracy: a hypothetical player 1 with $\mathcal{G}_1=\{g,h\}$,
$M_g\setminus\{1\}=M_h\setminus\{1\}=\{2,3\}$, $\phi_{12}=0.4$,
$\phi_{13}=-0.6$, gives $c_{1,g}=c_{1,h}=-0.2$ identically, so
$w_{1,g}=w_{1,h}=0.5$ for every $\beta$ (Lemma~\ref{lem:recoverability}).
This is not a validity failure --- Requirement~\ref{req:validity} still
holds exactly --- but an honest report that no evidence distinguishes the
two labels: both groups receive the same two co-members contributing the
same combined signal, so the construction correctly reports it has nothing
to go on, rather than inventing a preference. A softened version ($M_g\setminus\{1\}=\{2,3,4\}$,
$\phi_{14}=0.05$) gives $w_{1,g}\approx0.5125$, $w_{1,h}\approx0.4875$:
near-degenerate inputs produce near-uniform, not discontinuous, output ---
the construction's behavior changes smoothly as the data moves toward or
away from the degenerate case, never jumping abruptly.

\begin{table}[h]
\centering
\begin{tabular}{@{}lcc@{}}
\toprule
Case & $w_{1,g}$ & $w_{1,h}$ \\
\midrule
Exact degeneracy & 0.5000 & 0.5000 \\
Near-degeneracy ($+\phi_{14}=0.05$) & 0.5125 & 0.4875 \\
\bottomrule
\end{tabular}
\caption{Graceful degradation: valid but uninformative at the boundary,
near-uniform (not discontinuous) near it.}
\label{tab:degenerate}
\end{table}
%%%%%%%%%%%%%%%%%%%%%%%%%%%%%%%%%%%%%%%%%%%%%%%%%%%%%%%%%%%%%%%%%%%%%%%%
\section{Limitations and Future Work}
\label{sec:discussion}

This section states where Theorem~\ref{thm:uniqueness} does and does not apply, rather than leaving a reader to discover its boundaries unannounced.

\textbf{Richness and eligibility.} Theorem~\ref{thm:uniqueness} requires a
player to be eligible: rich, via a witnessing group with at least two co-members and at least one exclusive co-member, and belonging to at least two known groups. This fails for
players whose every known group is anchored by at most one other member (Remark~\ref{rem:richness}), or whose only witnessing group is their sole group (Remark~\ref{rem:degeneracy}). The construction remains valid
throughout in either case (Proposition~\ref{prop:validity} does not depend
on eligibility), but is no longer uniquely forced. This is an honest
boundary of the characterization, not a silent failure: a non-eligible
player still gets a perfectly usable weight out of
formula~\eqref{eq:softmax}, as Section~\ref{sec:empirical}'s player~2
shows directly; what is lost is only the guarantee that no other rule
could have produced a different, equally valid answer for that player.

\textbf{Independence of the remaining requirements.} Section~\ref{sec:uniqueness}
gave explicit counterexamples showing four of the ten requirements are
genuinely load-bearing --- drop any one of them and some other, invalid
rule becomes admissible. We have not established that
Requirements~\ref{req:continuity}, \ref{req:validity}, or
\ref{req:iia} are individually non-redundant given the rest: it remains
possible, though we have no evidence either way, that one of these three
could in principle be derived from the others rather than needing to be
assumed on its own.

\textbf{Scope.} Consistent with the axiomatic-characterization tradition
this paper follows \citep{shapley1953,young1985,owen1977}, our results
establish what construction is forced, not how to compute the underlying
pairwise interaction terms at scale, nor how those terms should be
estimated when $v$ itself must be estimated rather than given exactly:
throughout this paper, $\phi_{ij}$ is treated as a known input, not
something to be learned or approximated from data. Extending this
characterization to higher-order Harsanyi interactions (beyond the pairwise
terms this paper relies on), to non-eligible overlap structures via a
weaker, non-unique but still principled rule, and to settings where group
structure is itself learned rather than exogenously fixed, are natural
directions for future work.
%%%%%%%%%%%%%%%%%%%%%%%%%%%%%%%%%%%%%%%%%%%%%%%%%%%%%%%%%%%%%%%%%%%%%%%%
\section{Conclusion}
\label{sec:conclusion}

We formalized membership attribution: distributing a player's known,
overlapping group affiliations --- a legislator's standing across
committees, or a firm division's standing across the business units it
reports to --- into normalized weights derived from a cooperative game's
own pairwise interaction structure. This question falls outside all three
of cooperative game theory's existing tools: payoff allocation, interaction
attribution, and coalition formation and stability alike, each of which
either assumes a player sits in exactly one group or does not take group
structure as given at all. We proved the direct, ratio-based approach to
this question is impossible on any mixed-sign profile --- the ordinary
case, not an edge case --- and proposed a restricted-domain softmax
construction in its place. We then proved this construction is the unique
solution satisfying five requirements, given a precisely defined richness
condition, with a sixth requirement pinning down the sign of its
temperature parameter and a seventh, shown independent via a fully
verified witness, forcing a single shared temperature across players. An
exact numerical example confirmed these properties hold in practice and
showed the construction degrades gracefully, rather than breaking, at the
one boundary the theorem itself identifies. The result is a closed-form
rule, forced by these requirements rather than chosen by hand, for
distributing one player's standing across an exogenously given set of
overlapping groups --- applicable wherever cooperative-game players carry
known, overlapping tags: legislative committees, interdisciplinary
research groups, or multi-unit organizations.

\bibliographystyle{plainnat}
\bibliography{references.bib}

\end{document}